\documentclass[%
 reprint,
amsmath,amssymb,
aps,
pra,
]{revtex4-2}

\usepackage{dcolumn}
\usepackage{bm}

\usepackage{graphicx}
\usepackage{subfigure}
\usepackage{graphicx,hyperref}
\expandafter\let\csname equation*\endcsname\relax
\expandafter\let\csname endequation*\endcsname\relax
\usepackage{amsmath}
\usepackage{amssymb}
\usepackage{multirow, makecell, comment} 
\newcommand{\fla}[1]{\begin{flalign}#1\end{flalign}}
\usepackage{braket}
\usepackage{lineno}
\usepackage{lipsum}
\usepackage{xcolor}
\usepackage{soul}
\usepackage{array}

\begin{document}

\preprint{APS/123-QED}

\title{Optical Quantum Memory Processing for Quantum Hashing}

\author {S. A. Moiseev}
\email{samoi@yandex.ru}.
\author{M. M. Minnegaliev}

\affiliation{Kazan Quantum Center, Kazan National Research Technical University n.a. A.N. Tupolev-KAI, K. Marx, 10,  Kazan, 420111, Russia}

\date{\today}

\begin{abstract}

Sender authentication is an important requirement for secure quantum communication networks.
We propose programmable  photon-echo quantum memory  processing of a  multi-time-bin photonic qudit for implementing a quantum-hash-based verification protocol.  
In this approach, using QM, sender (Alice) generates a quantum hash state by transforming a single stored photonic wave packet into the programmable photonic qudit.
Bob's memory coherently maps the received time-bin photonic qudit into a single output mode according to phase encoding.
Our scheme relies on a time-resolved coherent detection method introduced here, in which the  coherent responses from individual time-bins are sequentially accumulated within Bob's spin coherence prior to the final sender verification.
The key advantage of this coherent detection for quantum hashing is shown, and its potential applications are also discussed.
The performed analysis demonstrates that experimentally verified parameters of contemporary optical quantum memories enable a proof of principle realization of the proposed architecture.
These results reveal an additional role for quantum memories, serving not only as synchronization and repeater resources for long-distance quantum communication, but also as active elements for quantum hashing in conventional quantum key distribution  systems.





\end{abstract}


\maketitle


\section{Introduction}

Optical quantum communications, driven primarily by quantum key distribution (QKD) technology, have achieved significant success in quantum information science by enabling a high level of information security \cite{gisin2002quantum,pan2024evolution}.
The use of fundamental principles of quantum physics underlies the protocols being developed, which do not allow the use of classical devices in carrying out a number of basic operations with quantum information carriers \cite{nielsen2010quantum}. 
At the same time, the emerging practical problems with the implementation of the proposed quantum approaches and algorithms initiate the need for their further development.
In particular, improving long-distance QKD requires quantum-repeaters based on multimode and multi-qubit optical quantum memory (QM) \cite{sangouard2011quantum,Azuma2023,zhu2025remote, akhmedzhanov2025quantum}. 
Such QMs are used not only to store quantum states, but also to synchronize probabilistic processes and enable the coordinated operation of neighboring links in a quantum network. 
Recently, an optical QM has also been employed in a  storage-based implementation of Wiesner's unforgeable quantum-money protocol \cite{mamann2025quantum}.
This result highlights the potential of optical QMs as operational elements of quantum-information protocols, beyond their conventional roles in storage and synchronization.
A central objective is therefore the development of QMs capable of reliably storing and manipulating photonic states with multiple temporal modes. 
In this context, particular attention has been devoted to the storage of time-bin photonic qubits \cite{Tittel-LPR-2010,sangouard2011quantum,moiseev2025optical,
akhmedzhanov2025quantum,tittel2025quantum,2022-NPJQI-Ortu,gisin2007storage,
moiseev2010temporal,2021-NatComm-One_hour_AFC,ortu2022storage}.

However, information security challenges remain that must be addressed even when using QKD systems to connect adjacent nodes (Alice and Bob) in a quantum network. 
Specifically, establishing a key between Alice and Bob requires authenticating the identities of the parties that generated the shared key. 
In classical communications, this is achieved by using hash functions and corresponding (hash) algorithms. 
A cryptographic hash function is a deterministic function 
$h: {\mathcal X} \to {\mathcal Y}$ which  converts the input data (the message) $x\in{\mathcal X}$ into a short hash (the hash of $x$) $h(x)=y\in{\mathcal Y}$,which identifies the original data with high probability. 
Cryptographic hash functions are typically required to provide preimage resistance (one-wayness), collision resistance, and strong sensitivity of the
hash value to changes in the input. 
Preimage resistance means that it is computationally infeasible to recover an input from its hash value.
Collision resistance means that it is computationally infeasible to find two distinct inputs $x\neq x'$ such that $h(x)=h(x')$.
The latter property is commonly associated with the avalanche effect, whereby a small change in the input produces a substantial change in the hash value.
This basic technology is used as the fundamental primitive for ensuring data integrity, generating digital signatures, content fingerprinting, and various other and modern security protocols \cite{Diffie1976,Kaye2006,security2010,Hash_blochain}.
Researchers continue to actively develop increasingly sophisticated methods to implement this fundamental cryptographic protocol \cite{mehmood2024advances}.

In quantum information technologies, hash functions have also become an integral component of QKD systems being developed  \cite{bennett1995generalized}.
Currently, for the experimental implementation of quantum hashing, it is proposed to use quantum walk algorithms \cite{hou2023quantum,giri2026quantum} and Gaussian boson sampling on a photonic quantum computer \cite{hatanaka2024quantum}.
The quantum hashing approach proposed in \cite{ablayev2014cryptographic,ablayev2016balanced,vasiliev2023constant} appears to be particularly suitable for quantum communication applications.
In this approach, the hash function encodes a classical input state into a quantum state of quantum carriers  so that the compression of each input words of length m to image of $log(m)$ length.
The generated quantum database state with a high probability satisfies to the condition of applying quantum amplification search and  using this algorithm one can exponentially reduce the number of qubits required.
The search for effective ways to experimentally implement quantum hashing is an urgent task in the development of this quantum technology.
Several types of quantum states have been proposed for encoding classical input data, see recent review \cite{ablayev2025theory}. 
Below we will focus on the use of one of them, which has been implemented experimentally in recent work \cite{akat2022multiqudit}, where the quantum hash encoding uses single-photon high-dimensional states (qudits).

Using significant progress in the experimental development of optical QM (see recent reviews \cite{Chaneliere2018,guo2023rare,Moiseev_fp2025} and section III below), in this work we develop the photon echo QM protocol \cite{moiseev2001complete,moiseev2004photon}in a quantum processing mode for  implementation of  quantum hash generation and phase-sensitive verification.
The principle scheme of the proposed protocol contains several stages.
Using QM device, Alice prepares an encoded multi-time-bin single-photon qudit to serve as the quantum hash state.
This quantum hash is directed to Bob, where it is stored in his (second) QM cell. 
By applying a known password-dependent sequence of control pulses during the storage process, Bob decodes the quantum hash, resulting in mapping all the photonic time bins onto a single spin wave in its QM. 
He then retrieves a single-photon wave packet from the QM at a target time with high efficiency, which serves as  verification signal associated with the encoded quantum hash. 
Next, we show the existing experimental possibilities of implementing the proposed protocol, using the latest advances in the implementation of optical QM on crystals with rare earth ions.
\section{Multimode quantum memory processing for quantum hashing}
\subsection{Theoretical model}

Let's consider a basic three-level scheme for the formation of QM based on photon echo \cite{moiseev2001complete,Tittel-LPR-2010, Chaneliere2018,Moiseev_fp2025}. 
This scheme uses an ensemble of atoms with large inhomogeneous broadening at optical $\ket{1}\leftrightarrow\ket{3}$ transition with linewidth  $\Delta_{in}$  exceeding the spectral width of the signal radiation $\delta\omega_s$ (see Fig.1) 
Below we will consider the use of various options for inhomogeneous broadening. 
Initially, all $N$ atoms are prepared in the ground state
$\ket{\Psi_a(t\rightarrow-\infty)}=
\prod_{j=1}^{N}\otimes \ket{1_{j}}$.
A signal light field 
is also prepared in the state $\ket{\Psi_{f}(t\rightarrow-\infty)}$, and it is launched into the atomic system at a given moment of time $t=0$. 
We assume that the light pulse is weak, containing on average a small number of photons $n_{ph}\ll N$.
We will be interested in the case when the field pulses propagate along  $\overrightarrow{\textbf{z}}$ direction and described by the wave functions of single-photon: 

\fla{
\ket{\Psi_{f}(t\rightarrow -\infty)}=&   \ket{\Psi_{ph}(t\rightarrow -\infty)}
\nonumber \\
=&\int dk f(k-k_s;t) \hat{a}^{\dagger} (k)\ket{\emptyset},
\label{photon}
}
where $f(k-k_s;t)$   is non-zero inside the spectral range $\delta \omega_s$ with a maximum at $k=k_s$ and  $\int dk |f(k-k_s;t)|^2=1$, 
$\delta\omega_s$ is a spectral width of wave function $|f(k-k_s;t)|$,
$[\hat{a}_s(k), \hat{a}^{\dagger}_s(k')]=\delta(k-k')$, 
 $\ket{\emptyset}$ is a vacuum state of all the field modes,
$k_s=\omega_{s}/c$.

\begin{figure}
	\centering
		\includegraphics[scale=0.75]{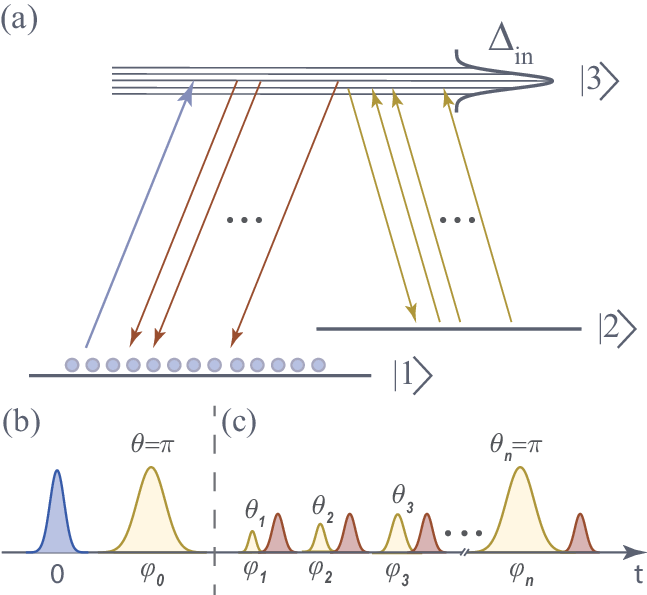}
\caption{(a) Energy level scheme and quantum transitions during atom-light interaction.
(b) Storage stage: Interaction with the signal pulse (blue) on the $\ket{1}\leftrightarrow\ket{3}$ transition, followed by a resonant writing $\pi$ control pulse on the $\ket{2}\leftrightarrow\ket{3}$ transition.
(c) Hash state generation: 
Application of a train of readout laser pulses (yellow) with  pulse areas 
$\theta_s$ and phases $\varphi_s$ controlled by Alisa, resulting in a cascade of photon-echo signal (brown) emissions with engineered amplitudes and phases. 
}
	\label{sequence-generation}
\end{figure}

For further, it is also convenient to use the description of the photon wave function using the photon creation $\hat{a}^{\dagger} (z)$ (and annihilation $\hat{a}(z)$) operator at a point with the coordinate $z$ (satisfying the  commutation relations $[\hat{a} (z),\hat{a}^{\dagger} (z')]=\delta(z-z')$):

\fla{
&\ket{\Psi_{ph}(t\rightarrow -\infty)}
= \int_{-\infty}^{\infty} dz \mathcal{A}\bigl(t-z/c \bigr) \hat{a}^{\dagger} (z)\ket{\emptyset},
\label{photon_2}
\\ 
\mathcal{A}\bigl(u \bigr)&=
\frac {e^{-i\omega_{s} u}}{\sqrt{2\pi}} 
\int d\kappa f_0(\kappa) e^{ic \kappa u}
=e^{-i\omega_{s} u}\mathcal{F}_{in}\Bigl({\frac{u}{\delta t_s}}\Bigr),
\label{photon_2b}
}
where
$\mathcal{A}\bigl(u \bigr)$ indicates forward propagation of photonic wave packet having a time duration $\delta t_s=\delta\omega_s^{-1}$,   
$\hat{a}^{\dagger} (z)=\frac {1}{\sqrt{2\pi}}\int dk e^{-ikz} \hat{a}^{\dagger} (k) $, $\hat{a}(z)=[\hat{a}^{\dagger} (z)]^{\dagger}$.
We also used 
$f(k-k_s;t\rightarrow- \infty)=f_0(k-k_s)e^{-i\omega_k t}=f_0(\kappa)e^{-i(\omega_s+c \kappa) t}$ 
and  
$f(k-k_s;t\rightarrow- \infty)
e^{ikz} = e^{-i\omega_s (t-z/c)}
f_0(\kappa)e^{-ic\kappa (t-z/c)}$.

A complete transfer of the quantum state of the input light pulse into the atomic system is realized in two cases. 
The first of them uses an atomic medium with an optically thick atomic transition $\alpha L\gg1$,
where $\alpha$ is a resonant absorption coefficient, $L$ linear size of the atomic medium.   
In the second case, the atoms are placed in an optical resonator, the frequency of which coincides with the frequency of the optical atomic transition. 
In this case, it is possible to use atomic ensembles with an optically thin resonant transition, but satisfying the impedance matching condition of  the resonant interaction \cite{Moiseev2010cavity,Afzelius-PRA-2010}.
Below, for the sake of simplicity, we will consider only one (the first) of these two possible QM schemes.

We will solve system of Heizenberg  equations for the operators of signal  light field and coherences of three-level atoms: 


\fla{
&\left( \frac{\partial}{\partial t} + v_g \frac{\partial}{\partial z} \right) \hat{b}_s(z,t) =  
-2\pi i \sum_{j=1}^{N} g_{s} \hat{S}_{13}^{j}(t)\delta(z-z_j),
\label{eq_b_s}
\\
& \frac{\partial \hat{S}_{13}^{j}(t) }{\partial t} = - i\Delta_{31}^j \hat{S}_{13}^{j}(t) 
-i g_s \hat{b}_s(z_j,t)-i\Omega(t)e^{-i\varphi} \hat{S}_{12}^{j}(t)  ,
\label{eq_S_13}
\\
& \frac{\partial \hat{S}_{12}^{j}(t) }{\partial t} = 
- i\delta_{21}^j \hat{S}_{12}^{j}(t) 
-i \Omega (t) ^{i\varphi} \hat{S}_{13}(t),
\label{eq_S_12}
}
where we assume resonant interaction $\omega_s=\omega_{31}$ of signal fields with atomic transition $\ket{1}\leftrightarrow\ket{3}$  
with its initial quantum state 
$\ket{\Psi_{in}(-\infty)}=\ket{\Psi_a(-\infty)}\otimes
\ket{\Psi_{ph}(-\infty)}$,
we have also introduced slowly varying operators of atomic coherences 
$\hat{S}^{j}_{13}(t)= e^{i\omega_{31} t  - i k_s z_j} \hat{P}^{j}_{13}(t)$,
$\hat{S}^{j}_{12}(t)=
e^{i(\omega_{21} t-(k_s-k_c)z_j}\hat{P}^{j}_{12}(t)$, 
($\hat{P}^{j}_{nm}=\ket{n}_{jj}\bra{m}$),
and light field
$\hat{b}_s(t,z)= e^{i\omega_s t - i k_s z}\hat{a}_s (t,z)$, ($k_c=\omega_{32}/v_g$,
$v_g$ is a group velocity), 
the field operators satisfy the commutation relation 
$[\hat{b}_s(t,z), \hat{b}^{\dagger}_s(t,z')]=2\pi \delta(z-z')$, 
$\hat{a}_s (t,z)=\int dk  \hat{a} (k,t)e^{ikz}$; 
$\Delta_{31}^j$ and $\delta_{21}^j$ are frequency offsets on the atomic transitions $\ket{1}\leftrightarrow \ket{3}$ and $\ket{1}\leftrightarrow \ket{2}$;
$g_s$ and $\Omega(t)$ are the Rabi frequencies for coupling of $j$-th atom with signal field and external classical fields on atomic transitions $\ket{1}\leftrightarrow \ket{3}$ and $\ket{2}\leftrightarrow \ket{3}$ (see Fig.1), $\varphi$ is a constant phase of the control laser field.
Classical light pulses are used for controlling atomic coherence.

For simplicity, we assume a sufficiently small frequency splitting between the two ground levels $n_{ref} \omega_{21} (c \alpha)^{-1} \ll 1$ ($\alpha$ is an resonant absorption coefficient on atomic transition $\ket{1}\leftrightarrow\ket{3}$). 
Otherwise, all the atom should be prepared on second lower sublevel  and suitable propagation directions of the subsequent laser control pulses must be selected.
At first consideration,  the lower spin transition $\ket{1}\leftrightarrow \ket{2}$ is assumed to be homogeneously broadened ($\delta_{21}^J=0$) with a long spin coherence time $T_2\gg \delta\omega_s^{-1}$. 
After studying the proposed protocol, we will discuss ways to account for the effect of an inhomogeneous broadened spin transition.

\subsection{Quantum storage stage}

Taking into account that a signal quantum light pulse is launched in the atomic medium at time $t=0$, we solve Heisenberg Eqs. \eqref{eq_b_s}-\eqref{eq_S_12}  for light and atomic operators.
Given the lack of a controlling laser field at the stages of signal pulse absorption, we find the following solutions of Eqs. \eqref{eq_b_s},\eqref{eq_S_13}:

\fla{
\hat{b}_s(t,z) = & 
\hat{b}_{s,0}(t-z/v_g)e^{-\alpha z/2}-i\frac{2\pi g_s}{v_g} \sum_{j=1}^N e^{-i\Delta_{31}^j t}\cdot
\nonumber \\
& \int_{0}^z dz' e^{-\alpha (z-z')/2} \hat{S}_{13,0}^j\delta(z'-z_j),
\label{signal_field}
\\
\hat{S}^{j}_{13}(t)= &\hat{S}^{j}_{13,0}
e^{-i\Delta_{31}^{j}t}
-ig_s e^{-i\Delta_{31}^{j}t-\alpha z_j/2}\tilde{b}_{s,0}(\Delta_{31}^{j}),
\label{signal_atoms}
}
where
$\tilde{b}_{s,0}(\Delta_{31}^j) =\int_{-\infty}^{\infty} dt e^{i\Delta_{31}^jt}\hat{b}_{s,0}(t)$ 
is a Fourier component of the input signal, $\hat{S}_{13}^j\equiv\hat{S}_{13}^j(-\infty)$.

The solutions \eqref{signal_field},\eqref{signal_atoms} are valid  when the spectral width $\delta\omega_s\sim \delta t_s^{-1}$ ($\delta t_s$ is temporal duration of the signal pulse) of the signal pulse is much smaller than the inhomogeneous line broadening $\delta\omega_s\ll\Delta_{in}$.

Taking into account the initial state of the atoms ($\hat{S}_{13}^j\ket{\Psi_{in}(-\infty)} =0$), in the future we will omit the first term in Eq. \eqref{signal_atoms} and $\hat{S}_{13,0}^j\delta(z'-z_j)$ in Eq.\eqref{signal_field}.
Assuming also the high optical density of the atomic medium $\alpha L\gg 1$ and using the solution \eqref{signal_field},\eqref{signal_atoms} we find that the signal field is completely absorbed in the medium causing the appearance of atomic coherence 
$\hat{S}^{j}_{13}(t)= 
-ig_s e^{-i\Delta_{31}^{j}t-\alpha z_j/2}\tilde{b}_{s,0}(\Delta_{31}^{j})$.
To preserve the quantum state of the signal field for a longer time and to conveniently control the quantum coherence of atoms, we launch first control laser pulse at time $t=t_0\gg \delta t_s$  with pulse area $\int dt \Omega_0(t) dt =\pi$ and phase $\varphi_0$.
Herein, we assume that the  laser pulse propagates in the same direction as the signal pulse (see Fig.1) leading to the 
transfer of all optical coherence to the spin coherence:

\fla{
\hat{S}^{j}_{12}(t_1>\tau)= 
-g_s e^{-t_{1,0}/T_2}e^{i(\varphi_0-\Delta_{31}^{j}t_0)-\alpha z_j/2}\tilde{b}_{s,0}(\Delta_{31}^{j}).
\label{long-lived_coherence}
}


As it is seen in Eq. \eqref{signal_atoms}, due to the large inhomogeneous broadening of the optical transition, macroscopic spin coherence is suppressed when $\delta\omega_s t_0\gg 1 $ due to the dephasing of atoms after the signal pulse absorption.  
This state turns out to be frozen under the weak influence of inhomogeneous broadening of the spin transition.
In the solution \eqref{signal_atoms}, we took into account the effect of decoherence effects and the introduction of a relaxation factor $e^{-t_0/T_2}$ with phase relaxation time $T_2$ determined by the magnetic interactions of the electron  and nuclear spins of an atom with the environment.
This influence is significant when considering quantum evolution over sufficiently long time intervals.
Using the atomic coherence stored in the medium \eqref{long-lived_coherence}, we will further be interested in the procedure for preparing the quantum hashing state by taking into account two initial quantum states $\ket{\Psi_{ph}(-\infty)}$ of the input signal field.

\subsection{Quantum hashing state generation}

The photon echo technique of QM allows for signal retrieval based on the rephasing of dipoles within an inhomogeneously broadened ensemble, regardless of whether this broadening is intrinsic to the material or artificially induced \cite{moiseev2001complete,Moiseev_fp2025}.
The optical coherence excited in these materials can be controlled  via appropriate QM protocols, including, for example, CRIB/GEM \cite{moiseev2001complete,alexander2006photon}, AFC \cite{deRiedmatten2008}, ROSE \cite{damon2011revival}, NLPE \cite{ma2021elimination} and PLMC \cite{moiseev2025_PLMC}.
Here we employ the AFC protocol by incorporating recent experimental advances in its realization \cite{deRiedmatten2008,2022-NPJQI-Ortu,guo2023rare,2026-PRL-DD_AFC_PrYSO}, where inhomogeneous broadening consist of the periodic structure of narrow resonant lines \cite{dubetskii1985echos}  $\Delta_j \in n_j \Delta_{0}$, where $n_j$ of $j$-th atom: $n_j=0, \pm 1, \pm 2, ...$, $\Delta_{0}\gg T_2^{-1}$. 
In this case, we can take into account the finite spectral width $\delta<\Delta_0$. 

In the procedure for preparing the quantum hashing state we use  retrieving the sequence of echo pulses  by exposing the atomic medium on transition $|2\rangle \leftrightarrow |3\rangle$   to a sequence of $m$ laser pulses with preset parameters \cite{moiseev2004photon} (see Fig.2).   
Consider step by step reading the first and subsequent retrieved echo signals, which is described by a system of equations for the new slowly varying field $\hat{b}_e(z,t)$ and  operators of atomic coherence $\hat{R}^{j}_{13}(t)$:

\fla{
&\left( \frac{\partial}{\partial t} - v_g \frac{\partial}{\partial z} \right) \hat{b}_e(z,t) =  
-2\pi i \sum_{j=1}^{N} g_{s} \hat{R}_{13}^{j}(t)\delta(z-z_j),
\label{eq_b_s-r}
\\
& \frac{\partial \hat{R}_{13}^{j}(t) }{\partial t} = - i\Delta_{31}^j \hat{R}_{13}^{j}(t) 
-i g_s \hat{b}_e(z_j,t)
\nonumber \\
& -i\Omega_m(t) e^{-i\varphi_{m}} \hat{S}_{12}^{j}(t)  ,
\label{eq_S_13-r}
\\
& \frac{\partial \hat{S}_{12}^{j}(t) }{\partial t} = 
-i \Omega_m(t) e^{i\varphi_{m}}\hat{R}_{13}(t),
\label{eq_S_12-r}
}
where 
$\hat{R}^{j}_{13}(t)= e^{i\omega_{31} t  + i k_s z}\hat{P}^{j}_{13}(t)$,
$\Omega_m(t)$ and $\varphi_{m}$ are the controlled Rabi frequency and phase of $m$-th reading laser pulse ($m=1,2, ...,n$).

The first reading control laser pulse having carrier frequency $\omega_{32}$, phase $\varphi_1$ and pulse area $\theta_1\neq\pi$ is launched at time $t_1>t_0$ .
The pulse propagates in backward direction with wave vector $-\textbf{k}_c$ and excites $j$-th atom at $t_1-\frac{z_j}{v_g}$.
After the action of this pulse, the atomic coherences take the form: 

\fla{
\hat{S}^{j}_{12}(t_2>t_1)= &
-\cos(\frac{\theta_2}{2})
g_s \Gamma_{1,2}(t_{2,0}) 
\nonumber \\
& e^{i(\varphi_0-\Delta_{31}^{j}t_0)-\alpha z_j/2}\tilde{b}_{s,0}(\Delta_{31}^{j}),
\\
\hat{R}^{j}_{13}(t>t_2)= &
ig_s \sin(\frac{\theta_2}{2}) 
 e^{-t_{1,0}/T_2}\cdot
 \nonumber \\ 
 &e^{-i\varphi_{1,0}} e^{-i\Delta_{31}^{j}(t-t_{1,0}+\frac{z_j}{v_g})-
\alpha z_j/2}\tilde{b}_{s,0}(\Delta_{31}^{j}),
\label{R_13}
}
where $\varphi_{m,n}=\varphi_{m}-\varphi_{n}$, $t_{m,n}=t_m-t_n$,
$\Gamma(t_{1,0})=
e^{-t_{1,0}/T_2}$.

The excited optical coherence $\hat{S}^{j}_{13}(t>t_2)$ now acts as an initial condition for subsequent evolution of Eqs.\eqref{eq_b_s-r},\eqref{eq_S_13-r}.   
Taking into account the nature of the inhomogeneous line (see comments  broadening $\Delta_j= n_j \delta$, we obtain that the macroscopic optical coherence will phase at a time $t=t_{e,1}=t_{1,0}+\tau$ (where $\tau=2\pi/\Delta$ and $\Delta_j \tau=2\pi n_j$) leading to the generation of first echo signal.
Substituting the coherence $\hat{S}^{j}_{13}(t>t_2)$ in the    initial condition of Eqs.\eqref{eq_b_s-r},\eqref{eq_S_13-r} (with $\Omega(t)=0$), we obtain the following equation for the echo signal emitted by the medium:

\fla{
&\left( \frac{\partial}{ v_g\partial t} -  \frac{\partial}{\partial z} \right) \hat{b}_e(z,t) =  
-\frac{\alpha}{2} \hat{b}_e(z,t) 
\nonumber \\
&-2\pi i \sum_{j=1}^{N} \frac{g_{s}}{v_g} \hat{R}_{13}^{j}(t>t_2)\delta(z-z_j)=-\frac{\alpha}{2} \hat{b}_e(z,t)
\nonumber
\\
&+\alpha 
\sin(\frac{\theta_1}{2})
e^{-i\varphi_{1,0}} 
\Gamma_{1,2}(t_{1,0})
\Gamma_{1,3}(\tau)
\hat{b}_s(t+\frac{z}{v_g}-t_{e,1})
 e^{-\alpha z/2},
\label{Eq_echo_field}
}
and obtain its following solution:

\fla{
\hat{b}_e(z,t)=&\sin(\frac{\theta_1}{2}) \Gamma_{1,2}(t_{1,0}) \Gamma_{1,3}(\tau) e^{-i\varphi_{1,0}}  e^{\alpha z/2}\cdot
\nonumber \\
& (e^{-\alpha z}-e^{-\alpha L})
\hat{b}_s(t+\frac{z}{v_g}-t_{e,1}),
\label{echo_field}
}
where
$\Gamma_{1,3}(\tau=\frac{2\pi}{\Delta_0})\cong\exp\{-\frac{1}{2}(\frac{\pi\delta}{\Delta_{0}})^2\}$ 
describes phase relaxation of atomic coherence on the optical transition $\ket{1}\leftrightarrow \ket{3}$, which is determined 
by Gaussian shape of AFC comb \cite{afzelius2009multimode}. 
At the medium output ($z=0$) in the limit of high optical depth ($\alpha L \gg 1$), the first echo signal is given by:

\fla{
\hat{b}_{e,1}(0,t)=\sin(\frac{\theta_1}{2}) \Gamma_{1,2}(t_{1,0}) \Gamma_{1,3}(\tau) e^{-i\varphi_{1,0}} 
\hat{b}_s(t-t_{e,1}).
}

When the echo signal is emitted, we reproduce the retrieval of the second and subsequent echo signals by acting on the atomic system with similar laser pulses having pulse areas $\theta_1, \theta_2, ..., \theta_n $. 
Here we assume that only the last $n$-th laser pulse has a pulse area $\theta_n=\pi$.
Following similar calculations, we obtain the solution for the field operator of the $m$-th echo signal ($1 < m \leq n$):


\fla{
\hat{b}_{e,m}(0,t)=&
\cos(\frac{\theta_1}{2}) 
\cdot ...\cdot
\cos(\frac{\theta_{m-1}}{2}) 
\sin(\frac{\theta_{m}}{2})\cdot
\nonumber \\
&\Gamma_{1,2}(t_{m,0})
\Gamma_{1,3}(\tau)
e^{-i\varphi_{m,0}} 
\hat{b}_s(t-t_{e,m}),
}
where $t_{e,m}=t_{m,0}+\tau$ is a time of $m$-th echo pulse emission, $t_{m,0}$ - the arrival time of the $m$-th laser pulse, $\varphi_{m,0}$ - phase of m-th echo signal. 

After the last $n$-th laser pulse, there is a complete transfer of atomic coherence from transition $\ket{1}\leftrightarrow \ket{2}$  to the transition  $\ket{1}\leftrightarrow \ket{3}$  and then to the light field, and this ends the retrieval of the quantum state of the signal pulse.
It is worth noting that by assuming negligibly weak phase relaxation in Eqs.\eqref{Eq_echo_field},
\eqref{echo_field}, we find 100\% efficiency of the signal pulse retrieval  in a series of emitted photon echo signals.
In this case, the  variant of equal probability ($\frac{1}{N}$) of emission of $n$ photon echo signals is interesting, which is realized if the pulse areas of the controlling laser pulses satisfy the condition:
$\sin(\frac{\theta_1}{2})=\frac{1}{\sqrt{n}}$,
$\sin(\frac{\theta_2}{2})=\frac{1}{\sqrt{n-1}}$, ... ,
$\sin(\frac{\theta_{n-1}}{2})=\frac{1}{\sqrt{2}}$,
$\sin(\frac{\theta_n}{2})=1$.
In this case, we get:

\fla{
\hat{b}_{e,m}(0,t)=
\frac{1}{\sqrt{n}}
\Gamma_{1,2}(t_{m,0})
\Gamma_{1,3}(\tau)
e^{-i\varphi_{m,0}} 
\hat{b}_s(t-t_{e,m}).
\label{hash-fields}
}

In the case of \textit{a single-photon input wave packet} and weak relaxation ($\Gamma_{1,3}(\tau)\rightarrow 1$, $ \Gamma_{12}(t_{m,0})\rightarrow 1$), the solution for the system's wave function reads:

\fla{
\ket{\Psi_{ph}(\varphi,\infty)}=
\sum_{m=1}^n \frac{e^{-i\varphi_{m,0}}}{\sqrt{n}}\ket{\Psi_{ph,m}(t\rightarrow\infty)} \otimes
\ket{\Psi_{a}(-\infty)}, 
\label{hash-state-1}
}
which reflects the same probability ($\frac{1}{n}$) of input signal pulse  recovery in each echo pulse, where $\bra{\Psi_{ph,m}(t\rightarrow\infty)}\ket{\Psi_{ph,m'}(t\rightarrow\infty)}=\delta (m-m')$.

Since all solutions for field operators give the same average values for these operators which are expressed in terms of the same initial quantum state \eqref{photon}, \eqref{photon_2}, it turns out that all wave the function $\ket{\Psi_{ph,m}(\infty)}$ includes the phase $\varphi_{m,0}$ and time of occurrence $t_{e,m}$ of 
the m-th echo signal.
By introducing additional indexes into the phases in the hash wave functions ($\varphi_{m,0}\rightarrow \varphi_{[j;m]}(x$), bearing in mind the specifics of preparing the quantum hashing state, we have:

\fla{
\varphi_{[j;m]}(x)=\frac{2\pi s_{j,m} x}{q},
\label{phi-m,x}
}
where $j=1,...,p$, $x\in \{0,1,...q-1\}$, 
$q$ is the size of the input state space  
\cite{ablayev2014cryptographic,ablayev2016balanced,vasiliev2023constant}.

It is possible to control the phases of the echo signals by manipulating the phase of the spin coherence of the atomic ensemble, for instance, by applying Stark pulses to the crystal.
Stark-controlled photon-echo protocols have already been demonstrated in rare-earth-ion quantum memories \cite{2020-PRL-EuYSO,2021-PRB-Minnegaliev, 2024-PRA-Alqedra,2026-PRA-EuYSO-AFC+Stark}. 
For the present protocol, Stark control is relevant because the hash information is encoded in relative phases between temporal modes. 
If the electric field acts differently on the relevant spectral classes or during different temporal intervals, the induced phases can be incorporated into the programmable phases $\varphi_{j,m}(x)$. 
In this way, Stark control may serve as an additional phase and timing control mechanism for the generation and verification of the temporal-mode quantum hash.

For the considered photon echo signals emitted in the backward  $-\overrightarrow{\textbf{z}}$ direction to the signal pulse, we find the following expression for the $m$-th wave function of single photon  state
$\ket{\Psi_{ph,m}(t\rightarrow\infty)}$:

\fla{
\ket{\Psi_{ph,m}(t\rightarrow \infty)}&= \int dk f(k+k_s;t) 
 e^{i\omega_k t_{e,m}}
\hat{a}^{\dagger} (k)\ket{\emptyset}
\nonumber
\\ 
&= \int_{-\infty}^{\infty} dz \mathcal{A}(t+z/c-t_{e,m}) \hat{a}^{\dagger} (z)\ket{\emptyset}.
}

As it is seen in Eqs. (18),(20), in comparison with the input signal pulse  wave function \eqref{photon}, \eqref{photon_2}, the wave function of $m$-th echo signal (22) 
$\mathcal{A}(t+z/c-t_{e,m}) $ 
reproduces the temporal shape of the input wave packet, but acquires an additional phase $\varphi_{m,0}$ and time delay $t_{e,m}$, which are controlled by the parameters of two laser pulses $0$ and $m$.
Taking into account the presence of phase relaxation, we obtain (for short, replacing $t\rightarrow \pm \infty$ by $\pm\infty$ in the wave function arguments):

\fla{
\ket{\Psi_{ph}^j(\varphi,\infty)}& =
\frac{\Gamma_{1,3}(\tau)}{\sqrt{n}}\sum_{m=1}^n \Gamma_{1,2}(t_{m,0}) e^{-i\varphi_{[j;m]}(x)}\cdot
\nonumber \\
&\ket{\Psi_{ph,m}(\infty)} \otimes
\ket{\Psi_{a}(-\infty)}
 + C_a \ket{\Psi_a(\infty)}, 
 \label{hash-state-2}
}
where the wave function $\ket{\Psi_a(\infty)}$ describes the state of the atomic medium that arose as a result of relaxation of atomic coherence and was not converted into a signal by controlling laser pulses.

Thus, we see that repeated reading of the state of a single-photon signal recorded in QM makes it possible to create a quantum hash state.
This is realized by repeating the scheme described above, introducing the sequence of values $\{s_{j,1}, ..., s_{j,n}\}$ in phases $\varphi_{[j;m]}(x)$  characterized by its index $j\in\{1,...,p\}$.

\begin{figure}
	\centering
		\includegraphics[scale=0.75]{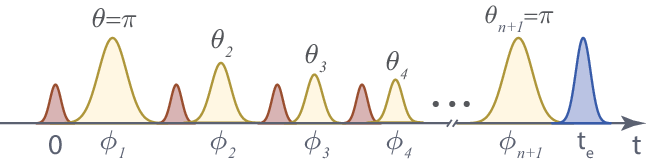}
	\caption{Hash state authorization stage: A multi-time bin photonic qudit (a sequence of brown temporal modes) are launched into Bob's QM cell. Subsequently, control laser pulses (yellow) with a tailored pulse areas and phases are applied after each time bin. Upon loading all bins, a readout $\pi$-pulse is applied, which retrieves the single-photon signal (blue) at time $t = t_e\equiv t_{e,n+1}$ with high probability (see Eq.\eqref{hash-echo}).
		}
	\label{sequence-bob}
\end{figure}

\subsection{Authentication protocol}

A crucial aspect of this protocol is the authentication of the quantum hash by Bob (the receiver). Specifically, the receiver must verify that their dataset matches the sender's with high fidelity.
Below, we demonstrate that our QM enables not only deterministic synthesis of hash state  \eqref{hash-state-2},  but also efficient sender authentication.

Fig. 2 illustrates the protocol for mapping the hash state into Bob's QM.
Bob employs a similar QM cell where incoming time-bin photonic states are loaded under dynamic control using additional laser pulses with specific pulse areas and phases. 
We show that the success of this process depends on the applied pulse phases, which thereby serve as unambiguous sender authentication.
Below, we provide a detailed description of this protocol.
We once more solve the system of Heisenberg equations \eqref{eq_b_s}-\eqref{eq_S_12} for the signal field operators and the atomic coherences of Bob's three-level atomic ensemble.
Initially, all atoms are assumed to be in the ground state, while the signal radiation is in the multi-time-bin single-photon states \eqref{hash-state-1} and \eqref{hash-state-2}, or,
in the more general case, the field operators are described by Eq. \eqref{hash-fields}.


\textbf{Mapping the first time-bin on long-lived atomic coherences.} 
Similar to the interaction and mapping of the input signal pulse in Alisa's QM (see Eq.
\eqref{signal_atoms}), the first photonic time-bin   is completely mapped in the Bob's QM, causing the appearance of atomic coherences $S_{13}^j(t)$
 by the time $t_1=\tau$: 

\fla{
\hat{S}^{j}_{13}(t_1=\tau)&= 
-ig_s\frac{1}{\sqrt{n}} 
e^{-i\varphi_{1,0}} 
e^{-i\Delta_{31}^{j}\tau-\alpha z_j/2}\tilde{b}_{s,0}(\Delta_{31}^{j})
\nonumber \\
&\equiv-i\frac{e^{-i\varphi_{1,0}} }{\sqrt{n}}\mathcal{B}(\tau)_j.
\label{eq_S_13-de2}
}

Next, the first short laser $\pi$-pulse will be applied, which fully transfers the  optical atomic coherence $\hat{S}^{j}_{13}(t)$ to long-lived spin coherence $\hat{S}^{j}_{12}(t)$:

\fla{
\hat{S}^{j}_{12}(t>\tau)= 
\hat{S}^{j}_{12}(\tau)=
-\frac{e^{i(\phi_{1}-\varphi_{1,0})}}{\sqrt{n}}\mathcal{B}_j(\tau), 
\label{eq_S_12-de1}
}
where $\phi_{1}$ is a phase of first control laser pulse on Bob side.

After this transfer, the second photonic time-bin is launched into Bob's QM cell.


\textbf{Mapping the second and subsequent photonic time-bins.}
The second photonic time-bin  is similarly absorbed by QM atoms, and after the same period of time $\tau$, we the optical coherence 

\fla{
\hat{S}_{13}^{j}(t_2)=-i\frac{e^{-i\varphi_{2,0}} }{\sqrt{n}}\mathcal{B}_j(t_2), 
\label{eq_S_13-de1}
}
as in Eq. \eqref{eq_S_13-de2}  is also excited in the atomic medium as when interacting with the first time-bin but with new phase $\varphi_{2,0}$. 
Next, the second short laser pulse with its phase  $\phi_{2}$ is applied, the action of which leads to the evolution of the atomic coherences described by the equations:

\fla{
& \frac{\partial \hat{S}_{13}^{j}(t) }{\partial t} = -i\Omega_2(t)e^{-i\varphi_{b,2}} \hat{S}_{12}^{j}(t)  ,
\label{eq_S_13-de}
\\
& \frac{\partial \hat{S}_{12}^{j}(t) }{\partial t} = 
-i \Omega_2 (t) ^{i\varphi_{b,2}} \hat{S}_{13}(t),
\label{eq_S_12-de}
}
with solutions

\fla{
&\hat{S}_{13}^{j}(t) =
\hat{S}_{13}^{j}(\tau)\cos(\frac{\theta_{b,2}}{2})
-i\sin(\frac{\theta_{b,2}}{2})
e^{-i\phi_{2}} \hat{S}_{12}^{j}(\tau),
\label{eq_S_13-n1}
\\
& \hat{S}_{12}^{j}(t) = 
\hat{S}_{12}^{j}(\tau)\cos(\frac{\theta_{b,2}}{2})
-i\sin(\frac{\theta_{b,2}}{2})
e^{i\phi_{2}}  \hat{S}_{13}(\tau).
\label{eq_S_12-n1}
}

Substituting the  pulse area of the second laser pulse  ($m=2$) determined by the ratio $\sin(\frac{\theta_{b,2}}{2})=\frac{1}{\sqrt{2}}$, we obtain:

\fla{
&\hat{S}_{13}^{j}(t) =
\frac{i}{\sqrt{2n}}
\bigl(-
e^{-i\varphi_{2,0}}+e^{-i(\phi_{2,1}+\varphi_{1,0})}\bigr) \mathcal{B}_j(t),
\label{eq_S_13-n2}
\\
& \hat{S}_{12}^{j}(t) =
-\frac{1}{\sqrt{2n}}
\bigl(e^{i(\phi_{1}-\varphi_{1,0})}+
e^{i(\phi_{2}-\varphi_{2,0})}\bigr) \mathcal{B}_j(t).
\label{eq_S_12-n2}
}

Taking into account the  solutions \eqref{eq_S_13-n2},\eqref{eq_S_12-n2}, we impose a condition on the phases of the controlling pulses
$\varphi_{2,0}=\phi_{2,1}+\varphi_{1,0}$ 
(or $\phi_{2}=\phi_{1}+\varphi_{2,1}$), 
which ensures $\hat{S}_{13}^{j}(t)=0$ and complete transfer of quantum coherence to spin states:

\fla{
\hat{S}_{12}^{j}(t) =
-\sqrt{\frac{2}{n}} e^{i(\phi_{1}-\varphi_{1,0})}
\mathcal{B}_j(t).
\label{eq_S_12-n3}
}

Comparing Eq.\eqref{eq_S_12-n3} and \eqref{eq_S_12-de1}, we see an increase in the coherence amplitude from magnitude $\sqrt{\frac{1}{n}}$  to magnitude $\sqrt{\frac{2}{n}}$ and a change in phase from $\phi_{1}-\varphi_{1,0}$ to $\phi_{2}-\varphi_{2,0}$. 
Bob can store the third photonic time-bin in a similar way by using the pulse area of the next controlling laser pulse satisfying the condition $\sin(\frac{\theta_{b,3}}{2})=\frac{1}{\sqrt{3}}$ and phase  $\phi_{3}=\phi_{1}+\varphi_{3,1}$. 
In this case, we again have a complete transfer of atomic coherence from the optical transition to the spin transition with the formation of the next spin coherence:

\fla{
\hat{S}_{12}^{j}(t) =
-\sqrt{\frac{3}{n}} e^{i(\phi_{1}-\varphi_{1,0})}
\mathcal{B}_j(t).
\label{eq_S_12-n4}
}

By storing subsequent time-bins  in a similar way, using pulse areas for $4,5,..,m$-th laser pulses ($m=2,3,...n$) with pulse areas $\sin(\frac{\theta_{b,m}}{2})=\frac{1}{\sqrt{m}}$, $\bigl(\cos(\frac{\theta_{b,m}}{2})=\sqrt{\frac{m-1}{m}}\bigr)$  and phases $\phi_{m}=\phi_{1}+\varphi_{m,1}$, 
we obtain after the action of the $m$-th laser pulse
   
\fla{
\hat{S}_{12}^{j}(t) =
-\sqrt{\frac{m}{n}} e^{i(\phi_{1}-\varphi_{1,0})}
\mathcal{B}_j(t),
\label{eq_S_12-fn}
}
and after $m=n$ we get 
$\hat{S}_{12}^{j}(t) =
 e^{i(\phi_{n}-\varphi_{n,0})}
\mathcal{B}_j$, which is equivalent to storage of input signal pulse, given by solution \eqref{signal_atoms}.
Thus, optical coherence arising from the absorption of each time-bin is transferred onto the spin coherence by the action of the subsequent laser pulses with appropriate pulse  areas $\theta_{b,m}$ and phases $\phi_m$.
When the condition $\phi_{m} = \phi_{1} + \varphi_{m,1}$ is satisfied, all partial spin coherences interfere constructively to form a total coherence.

Subsequently, by applying a resonant readout laser pulse with an area of $\pi$ and phase $\phi_{n+1}$, Bob retrieves the photon echo signal at a predetermined time $=t_{e,n+1}$. 
The amplitude of this echo is given by:


\fla{
\hat{b}_{e}(0,t)=
 e^{-i(\phi_{n+1,1}+\varphi_{1,0})}
\hat{b}_s(t-t_{e,n+1}).
\label{echo-1}
}
Thus, selecting precise pulse areas $\theta_m$ and phases $\varphi_m$ for all $n$ readout laser pulses isolates the target echo mode. In contrast, a phase mismatch suppresses this desired signal and generates additional parasitic echoes from preceding excitations.
In this case, using a single-photon input state ensures efficient retrieval of the photon with a probability close to 100\% exclusively in the echo signal \eqref{echo-1}, which enables unambiguous sender authentication.

Leaving a more detailed statistical analysis of spurious echo emission under pulse parameter deviations for future work, we briefly examine the impact of relaxation processes in the QM on the retrieval efficiency of the controlled photon echo signal.
\\


\subsection{Influence of relaxation effects}

By applying Eq. \eqref{hash-fields} for the first input  time-bin hash field and 
taking into account phase relation on optical and spin transitions, we get the following atomic coherence $S_{13}^j(t)$ in the Bob's QM:

\fla{
\hat{S}^{j}_{13}(t_1=\tau')=-i\frac{e^{-i\varphi_{1,0}} }{\sqrt{n}}
\Gamma_{1,2}(t_{1,0})
\Gamma_{1,3}(\tau+\tau')
\mathcal{B}_j(\tau') ,
\label{S_13-f2}
}
and for the spin coherence after the of first laser pulse 

\fla{
\hat{S}_{12}^{j}(t) =
-\sqrt{\frac{1}{n}} e^{i(\phi_{1}-\varphi_{1,0})}
\Gamma_{1,2}(t_{1,0})
\Gamma_{1,3}(\tau+\tau')
\mathcal{B}_j(t).
\label{eq_S_12-f}
}

Then similarly, we get optical coherence resulting from the absorption of the second input time-bin hash field

\fla{
\hat{S}^{j}_{13}(t)=-i\frac{e^{-i\varphi_{2,0}} }{\sqrt{n}}
\Gamma_{1,2}(t_{2,0})
\Gamma_{1,3}(\tau+\tau')
\mathcal{B}_j(t).
\label{signal_field_and_atoms}
}

Thus, we see that relaxation effects affect the optical  coherences in the same way, which will therefore interfere with the same amplitudes. Taking into account appropriate phase relations (see comments to Eqs. \eqref{eq_S_13-n2},\eqref{eq_S_12-n2}), we get:   

\fla{
\hat{S}_{12}^{j}(t) =
-\sqrt{\frac{2}{n}} e^{i(\phi_{1}-\varphi_{1,0})}
\Gamma_{1,2}(t_{2,0})
\Gamma_{1,3}(\tau+\tau')
\mathcal{B}_j(t).
\label{eq_S_12-f1}
}

Next, by induction, we similarly obtain spin coherence after transferring the $m$-th photon echo signal to spin levels:

\fla{
\hat{S}_{12}^{j}(t) =
-\sqrt{\frac{m}{n}} e^{i(\phi_{1}-\varphi_{1,0})}
\Gamma_{1,2}(t_{m,0})
\Gamma_{1,3}(\tau+\tau')
\mathcal{B}_j(t),
\label{eq_S_12-f2}
}
and after the last $n$-th laser pulse:

\fla{
\hat{S}_{12}^{j}(t) =
-
\Gamma_{1,2}(t_{n,0})
\Gamma_{1,3}(\tau+\tau')
e^{i(\phi_{1}-\varphi_{1,0})}
\mathcal{B}_j(t).
\label{eq_S_12-f3}
}

Thus, the effect of spin relaxation appears to be the same as that of the last photon echo signal. 
The same effect of phase relaxation is of great importance for the ideal interference of photon echo signals in spin coherence.

After the quantum hash is fully loaded, one can similarly find  the amplitude of the reconstructed single photon echo:

\fla{
\hat{b}_{e}(0,t)=&
\Gamma_{1,2}(t_{n+1,0})
\Gamma_{1,3}(2\tau)\cdot
\nonumber \\
& e^{-i(\phi_{n+1,1}+\varphi_{1,0})}
\hat{b}_s(t-t_{e,n+1}).
\label{hash-echo}
}

According to this solution, the efficiency $E$ of photon echo generation (and sender authorization, respectively) will be

\fla{
E=|\Gamma_{1,2}(t_{n+1,0})
\Gamma_{1,3}(2\tau)|^2.
}

Taking into account the exponential character of phase relaxation at optical and spin transitions, we get  

\fla{
E=
\exp\{-2\gamma_{21}t_{n+1,0}\} 
\exp\{-2 (\frac{\pi\delta}{\Delta_{0}})^2\}.
\label{efficiency}
}

As it seen in the solution \eqref{efficiency}, the optical and spin decoherence  processes occur twice in both Alice's and Bob's QM cells. 
Meanwhile, spin relaxation affects the protocol efficiency at longer times ($t_{n+1,0} \gg \frac{4\pi}{\Delta_0}$), which imposes stringent requirements on the spin coherence lifetime of the atomic ensemble.
This challenge is currently addressed by various techniques, including multi-pulse radio-frequency dynamical decoupling \cite{souza2011robust,wang2025nuclear}.


\section{Experimental feasibility}

The proposed QM hash protocol  is based on a three-level optical quantum-memory scheme. 
The input wave packet is first absorbed on the optical transition $|1\rangle\leftrightarrow|3\rangle$ and is subsequently transferred to the long-lived spin coherence $|1\rangle\leftrightarrow|2\rangle$ by a control pulse with an area close to $\pi$. 
The hash-generation stage then uses a sequence of programmable readout pulses with controlled areas and phases.
The pulse areas determine the amplitudes of the retrieved temporal modes, whereas the pulse phases determine the relative phases $\phi_{j,m}(x)$ that encode the quantum hash. 
The proposed  protocol can employ the AFC spin-wave technique with programmable optical control, as it is the most developed photon-echo-based QM approach
with programmable optical control, phase-preserving spin-wave storage, and  on-demand readout.

A particularly relevant experimental benchmark is provided by the recent demonstration of a cavity-enhanced spin-wave AFC memory in Pr$^{3+}$:Y$_2$SiO$_5$ crystal \cite{2025-PRL-FeldmanPrYSO}. 
Weak coherent pulses containing 0.42 photons per pulse were stored and retrieved with an efficiency of up to 40\% and a signal-to-noise ratio of approximately 14.
The experiment also demonstrated an excited-state to spin-state  transfer efficiency of about 94\% along with the  storage and retrieval of heralded single photons.
More recently, nonclassical storage of telecom-heralded single photons has been demonstrated in a AFC spin-wave memory combined with an XY4 dynamical decoupling spin-rephasing sequence \cite{2026-PRL-DD_AFC_PrYSO}. 
Long-lived AFC echoes were observed for storage times of up to approximately  3 ms in the classical regime, while nonclassical correlations between the heralding and retrieved photons were preserved for storage times up to 180 $\mu$s, with a measured cross-correlation of 4.6.
These experimentally demonstrated capabilities enable repeated read and write operations of the multi-qudit hash states in Alice's and Bob's QM cells.


Dynamical decoupling is particularly relevant to the proposed protocol because it can compensate the effect of inhomogeneous broadening of the hyperfine spin transition. 
Since the hash is encoded in the relative phases between temporal modes, maintaining spin-wave phase coherence is as critical as preserving retrieval efficiency.
The number of temporal modes is determined by the available spin-wave coherence time and the temporal separation between neighboring wave packets.
The temporal modes should satisfy $|t_m-t_{m'}|\gg\delta t_s$, while the total hash-generation interval should obey $ nT_{\rm slot}+T_{\rm ctrl}\lesssim T_{\rm sw}$.
The effect of dynamical decoupling on the temporal hash must therefore be characterized not only through the retrieved intensity, but also through the relative phases and the overlap matrix of the temporal modes.

 Eu$^{3+}$:Y$_2$SiO$_5$ crystal is attractive for long storage times owing to its exceptionally long optical and spin coherence times, although generally at the expense of lower efficiency and more demanding control requirements.
Recent work on an integrated impedance-matched cavity containing Eu$^{3+}$:Y$_2$SiO$_5$	 demonstrated an AFC storage efficiency of up to 70\%
for telecom-heralded single photons, together with the storage of 20 temporal modes with an average efficiency of 51 \%  \cite{2026-NatPhot-RecordAFC_EuYSO}. 
These results are particularly relevant to the proposed protocol because they combine high-efficiency AFC storage with temporal multimode operation, which is required for the generation and coherent analysis of the proposed temporal-mode quantum hash. 

The preservation and analysis of time-bin coherence in rare-earth-ion AFC memories have already been demonstrated. 
In particular, a spin-wave AFC memory in $^{151}$Eu$^{3+}$:Y$_2$SiO$_5$ stored six temporal modes and two time-bin qubits; the latter were stored for 20 ms with an average output-state fidelity of $(85\pm2)\%$ \cite{2022-NPJQI-Ortu}. 
Their phase-sensitive analysis was performed at readout by a composite control pulse, without an external unbalanced interferometer. 
This result is relevant to our protocol because Bob's memory can preserve, and subsequently interfere, the relative phases of different time bins.
In addition, one-hour coherent optical storage has previously been demonstrated in $^{151}$Eu$^{3+}$:Y$_2$SiO$_5$ using a ZEFOZ magnetic field and dynamical decoupling, with a time-bin interference fidelity of approximately 96.4 \% after one hour \cite{2021-NatComm-One_hour_AFC}.
Further improvement of the control-pulse transfer efficiency and suppression of control-induced noise could therefore make Eu$^{3+}$:Y$_2$SiO$_5$ an especially attractive platform for implementing the proposed QM-based hashing protocol, particularly in applications requiring long synchronization delays and coherent processing of multiple temporal modes.

Er$^{3+}$:Y$_2$SiO$_5$ is of particular interest for quantum-memory-based hashing because the relevant optical transitions lie in the telecommunication band. 
This eliminates the need for an optical frequency conversion stage between the quantum memory and standard fiber-based communication channels.
In addition, $^{167}$Er$^{3+}$:Y$_2$SiO$_5$ offers long-lived hyperfine states \cite{2017-Rancic-NaturePhysics, 2023-JETPLetters-Minnegaliev}, with coherence times between ground-state hyperfine levels exceeding one second \cite{2017-Rancic-NaturePhysics} under optimized magnetic-field and cryogenic conditions.
These properties make erbium-doped Y$_2$SiO$_5$ an attractive candidate for long-delay, telecom-compatible quantum-memory protocols. 
AFC memories based on Er$^{3+}$:Y$_2$SiO$_5$ have already demonstrated storage of telecom-band single-photon-level light \cite{2021-PRR-Stuart,2023-NatComm-ErYSO,2026-BullRAS-Gerasimov}, including integrated on-demand AFC operation controlled by electric-field-induced Stark control \cite{2021-Optica-Faraon-Craiciu,2022-PRL-ErYSO-Integr_Stark}.

At the same time, a complete spin-wave AFC implementation in Er$^{3+}$:Y$_2$SiO$_5$ remains less mature than the corresponding implementations in Pr$^{3+}$:Y$_2$SiO$_5$ and Eu$^{3+}$:Y$_2$SiO$_5$. 
In particular, further improvements are required in the optical-to-spin transfer efficiency, control-pulse noise suppression, and preservation of phase coherence during spin-wave storage via dynamical decoupling \cite{2026-arxiv-SWAFC, 2025-arxiv-SWAFC}.  
Consequently, Er$^{3+}$:Y$_2$SiO$_5$ should be regarded as a promising telecom-compatible platform for a future implementation of the proposed QM-based hash, rather than as the most immediate platform for the first proof-of-principle experiment.


Of interest is also the experimental implementation of the proposed QM hash protocol using a novel photon-echo-based QM approach \cite{moiseev2025_PLMC,moiseev2026deterministic}, where QM programming is carried out using richer macroscopic spin coherence control capabilities. The development of this variant will be addressed in a forthcoming study.

\section{Conclusion}


In this work, we demonstrate that programmable processing within a photon-echo QM enables the implementation of a quantum-hashing protocol \cite{ablayev2014cryptographic,ablayev2016balanced,vasiliev2023constant} with robust user authentication.
In the proposed scheme, by using QM cell, Alice encodes a single-photon quantum hash in multiple temporal modes and sends the resulting photonic qudit to Bob. 
Bob  uses a password-dependent train of control pulses, with prescribed pulse areas $\theta_m$ and phases $\phi_m$, to coherently map the temporal-mode components onto a common long-lived spin coherence of his QM. 
For the matching phase pattern, the accumulated coherence is retrieved as a photon echo at the target time $t=t_e$ (see Fig. 2), providing the phase-sensitive verification signal.
Because the responses associated with the $n$ temporal modes are coherently accumulated in the QM and retrieved in a single photon-echo event, the proposed scheme replaces the $n$ separate mode-resolved detection steps required in the mode-by-mode procedure of Ref.\cite{akat2022multiqudit} with one collective readout. 
In the dark-count-limited regime, the proposed QM hashing scheme can lead to the signal-to-noise ratio improvement by a  factor of $n$. 
Thus, one of the main result of the present work is also the proposed method of quantum measurement of multi-time-bin photonic state with coherent accumulation of photonic time-bins in total spin coherence of QM, which provides additional advantages in experimental implementation of quantum hashing.


It is also important to note that even a small deviation of Bob's input parameters from those used to encode the state affects the mapping of the temporal modes onto the spin coherence and changes the retrieved photon-echo signal. 
This sensitivity is consistent with the distinguishability required for quantum hashing, but it should not be interpreted as a proof of cryptographic collision resistance. 
A quantitative assessment would require an explicit analysis of the overlap between hash states corresponding to different inputs, together with the relevant noise and proposed detection model.

The present results establish the physical feasibility of the proposed QM-assisted hashing operation, but they do not demonstrate perfect sender authentication or constitute a complete cryptographic security proof. 
Further theoretical and experimental work should quantify the effects of finite retrieval efficiency, imperfect temporal-mode preparation, phase errors, control-pulse imperfections, detector noise, and background counts. 
In particular, a rigorous analysis is required to evaluate the false-acceptance and false-rejection probabilities, as well as the preimage and collision-resistance properties of the resulting quantum-hash implementation.
The study of these issues   is also a  separate task that deserves special attention.

The obtained results highlight a new role of quantum memories: beyond serving as synchronization and repeater resources for long-distance quantum communication, they can also act as quantum processors for quantum hashing in QKD systems. 
It is worth noting separately that the  generating and coherently accumulating photonic time-bins in spin coherence of QMs also serves as an effective method for controlling the spectrum of the signal pulse. 
During the generation stage, a periodic frequency comb of narrow lines is formed, with a spectral separation corresponding to the temporal interval between the photonic time-bins ($\Delta=2\pi/\tau$). 
Conversely, when the bins are merged into a single echo signal, all these narrow spectral lines combine into one broad, bell-shaped (Gaussian) line. 
This deterministic control of the optical spectrum may be useful for controlling  slowly propagating broadband light pulses, as demonstrated in \cite{bonarota2012atomic}, and can be useful for addressing other problems in coherent and quantum optics.

\section{acknowledgments}
The authors thank F.M. Ablaev and A.V. Vasiliev for insightful discussions regarding the quantum generalization of the hash protocol they are developing,
we also thank
K.I.Gerasimov, N.S.Perminov and L.R. Gilyazov for useful comments.
This research was supported by the Ministry of Science and Higher Education of the Russian Federation (FZSU-2026-0017).


\bibliography{apssamp.bib}

\end{document}